\documentclass[twocolumn]{aastex631}

\date{\small 1st of April 2026: Submitted to \textit{Acta Prima Aprilia}}

\begin{document}

\title{Antimatter Propulsion for Interstellar Travel via Positron Production from Potassium-40 Rich Biological Matter}


\author{C. Hall}
\affiliation{University of Georgia}
\email{cassandra.hall@uga.edu}  

\author
{L. N. H. P. Hall}
\affiliation{Center for Astropawysics \textbar Barkvard \& Sniffsonian}


\begin{abstract}
Antimatter-based propulsion is often cited as a physically plausible route to relativistic interstellar travel, and thus as a potential mechanism by which technologically advanced civilizations could expand throughout the Galaxy. Its difficulty may be central to the resolution of Fermi's paradox. Since the Universe should be teaming with advanced technological life, yet we see none, it may be that interstellar travel is simply too difficult. It has been suggested that the main difficulty with using antimatter as propulsion is its limited availability, assuming it must be artificially manufactured. In this paper, we demonstrate that naturally occurring potassium-40–rich biological matter (specifically bananas) is a promising, overlooked antimatter source for interstellar propulsion.
%
%
\end{abstract}

\section{Introduction}

Fermi's paradox is not a paradox, and it wasn't Fermi's \citep{forgan2019}. Nevertheless, we happily accept this example of the law of misnomy, and consider: if the Universe is teaming with intellectually advanced civilization, then where is everyone? One obvious barrier is the difficulty in achieving interstellar travel. 

Antimatter-based propulsion has long been recognized as one of the few physically plausible mechanisms for achieving relativistic velocities required for interstellar travel. Because matter-antimatter annihilation converts rest mass directly into energy at generally complete efficiency, even tiny quantities of antimatter can provide energies far exceeding those attainable with conventional. As a result, antimatter propulsion concepts have been explored in a variety of contexts, including rapid flyby missions, deceleration architectures, and kilogram-scale probes to nearby stellar systems.


A persistent and well-recognized limitation of antimatter propulsion concepts lies not in the annihilation physics itself, but in the sourcing, production, and storage of antimatter \citep{Jackson2022}. Engineered antimatter production remains technologically challenging, energetically expensive, and limited to extremely small quantities. That leaves us with a timely question: what role can naturally occurring antimatter sources play in propulsion applications for interstellar travel?


In this work, we explore the use of bananas as a naturally occurring source of antimatter for interstellar propulsion. Bananas are potassium-rich biological matter and therefore contain trace amounts of the naturally occurring radioactive isotope potassium-40 ($^{40}$K). A small fraction of $^{40}$K decays proceed via $\beta^{+}$ decay, producing positrons that subsequently annihilate with electrons, releasing their rest mass energy. 

In this paper, we quantify the energetic contribution of banana-derived positrons under conservative assumptions for a variety of probe masses, and explore compression factors required in each case.   This approach provides a lower bound on antimatter availability from naturally occurring sources, and offers a starting point for technologically ambitious antimatter production schemes.


\section{Method}

\subsection{Positron Production}

We estimate the total antimatter energy available from biological potassium via the $\beta^{+}$ decay channel of naturally occurring potassium-40 ($^{40}$K). A single unit of biological matter is taken to be a medium banana, assumed to contain $0.45\,\mathrm{g}$ of elemental potassium. The isotopic abundance of $^{40}$K is fixed at $0.0117\%$ of natural potassium. Only a small fraction of $^{40}$K decays proceed via positron emission; we adopt a $\beta^{+}$ branching ratio of $0.001\%$.

The total number of potassium atoms per banana is computed using the molar mass of potassium ($39.0983\,\mathrm{g\,mol^{-1}}$) and Avogadro’s number $N_A$. The number of $^{40}$K atoms is then obtained by applying the isotopic abundance, and the total positron yield follows from the assumed branching ratio. Each emitted positron is assumed to annihilate with an electron, releasing $1.022\,\mathrm{MeV}$ of energy per annihilation.


All positron annihilation energy is assumed to be  $100\%$ efficient, and the total amount of annhilation energy released per banana is $E_{\mathrm{banana}}$. 

\subsection{Relativistic Kinetic Energy Requirements}

The energy required to accelerate a probe to relativistic speeds is computed using the special-relativistic kinetic energy expression
\begin{equation}
E_{\mathrm{kin}} = (\gamma - 1)\, m c^{2},
\end{equation}
where $m$ is the probe mass, $c$ is the speed of light, and
\begin{equation}
\gamma = \frac{1}{\sqrt{1 - \beta^{2}}}
\end{equation}
with $\beta = v/c$.

For each probe mass and cruise velocity, the number of bananas required is computed by:
\begin{equation}
N_{\mathrm{banana}} = \frac{E_{\mathrm{kin}}}{E_{\mathrm{banana}}}.
\end{equation}

To account for deceleration at the destination, an idealized braking phase is included by doubling the total energy requirement, corresponding to an assumed symmetric acceleration and deceleration.

\subsection{Probe Scenarios}

We consider three probe scenarios:
\begin{enumerate}
    \item A macroscopic interstellar probe with mass $1000\,\mathrm{kg}$,
    \item A gram-scale probe representative of Starshot-like concepts ($1\,\mathrm{g}$),
    \item A speculative nanorobot with mass $1\,\mu\mathrm{g}$.
\end{enumerate}


\subsection{Interstellar Travel Times}

For reference, travel times to Alpha Centauri in Earth-years are computed assuming constant velocity:
\begin{equation}
t_{\alpha\,\mathrm{Cen}} = \frac{d}{\beta c},
\end{equation}
where $d = 4.37$ light-years. These times are shown as vertical markers on velocity-dependent plots. We do not consider relativistic time dilation effects in the probe frame.

\subsection{Storage Volume and Compression Factor Estimates}

To assess the feasibility of storing the required biological matter onboard each probe, a simplified volumetric analysis is performed. Each banana is assigned an uncompressed volume
\begin{equation}
V_{\mathrm{banana}} = 10^{-4}\,\mathrm{m^{3}}.
\end{equation}
Each probe is assigned a fixed available storage volume, chosen to scale plausibly with probe size.

For a given probe and cruise velocity, the total required storage volume is
\begin{equation}
V_{\mathrm{req}} = N_{\mathrm{banana}}\, V_{\mathrm{banana}}.
\end{equation}
The required compression factor is then defined as
\begin{equation}
C = \frac{V_{\mathrm{req}}}{V_{\mathrm{available}}}.
\end{equation}

Compression factors are computed for both acceleration-only and acceleration-plus-braking scenarios and plotted as a function of cruise velocity. No physical mechanism for achieving such compression is specified.



\section{Results}
\label{sec:results}

Figure~1 shows the main result of this work: the total number of bananas required to accelerate three representative payload classes to a given cruise velocity, shown for both acceleration-only (solid) and acceleration-plus-braking (dashed) mission profiles. The required banana count increases sharply with cruise velocity through the relativistic factor $\gamma(\beta)$, and scales linearly with payload mass. Consequently, the separation between the three payload classes is essentially constant in $\log N_{\rm banana}$ at fixed $\beta$, while all curves rise steeply as $\beta\rightarrow 1$.

\begin{figure}[h]
    \centering
    \includegraphics[width=\columnwidth]{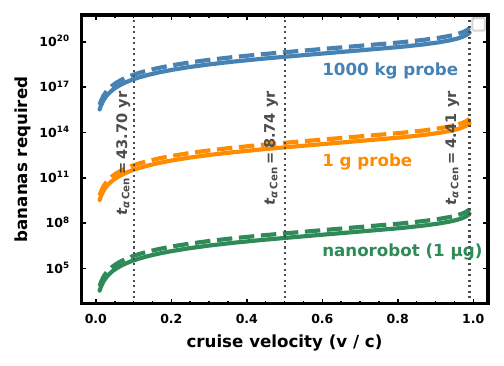}
    \caption{A demonstration of the number of bananas required to reach a given cruise velocity. }
    \label{fig:example}
\end{figure}

For a $1000$ kg probe, even modest relativistic cruise velocities imply extreme banana requirements. At $\beta=0.1$ (corresponding to an Earth-frame Alpha Centauri cruise time of $t_{\alpha\,{\rm Cen}} \approx 43.7$ yr), the banana requirement is already $\sim 10^{18}$ bananas for acceleration-only, rising by a factor of two for acceleration-plus-braking. At higher cruise velocities, the requirements become rapidly prohibitive: the curves approach $\gtrsim 10^{19}$ bananas as $\beta \rightarrow 1$, reflecting the divergence of $\gamma(\beta)$.

For a gram-scale probe (Starshot-like mass scale), the banana requirement is reduced by six orders of magnitude relative to the $1000$ kg case, but remains astronomically large for relativistic cruise. At $\beta=0.1$, the required banana count is $\sim 10^{12}$, again doubling if terminal braking is included. Only in the extreme nanorobot regime ($m \sim 10^{-9}$ kg) do the banana requirements become numerically modest: at $\beta=0.1$, the model predicts $\sim 10^{5}$ bananas for acceleration-only. 

Vertical markers in Figure~1 indicate representative cruise velocities and the corresponding Earth-frame transit times to Alpha Centauri, $t_{\alpha\,{\rm Cen}} \simeq 4.37/\beta$ yr. These markers highlight the fundamental trade-off identified by Figure~1: reducing mission duration by increasing cruise speed rapidly drives the banana requirement up. 

\begin{figure}[h]
    \centering
    \includegraphics[width=\columnwidth]{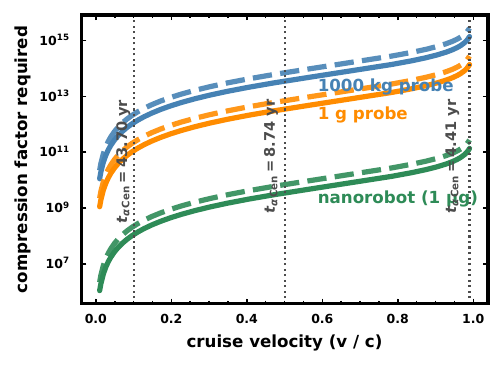}
    \caption{The compression factor required to fit the necessary number of bananas on board each probe.}
    \label{fig:example}
\end{figure}

Figure~2 translates the banana budgets of Figure~1 into a required ``compression factor,'' defined as
\begin{equation}
C \;=\; \frac{V_{\rm req}}{V_{\rm avail}}
\;=\; \frac{N_{\rm banana}\,V_{\rm banana}}{V_{\rm avail}},
\end{equation}
where we take a fiducial uncompressed banana volume $V_{\rm banana}=10^{-4}\,{\rm m}^{3}$ and impose fixed available storage volumes for each payload class (e.g., $V_{\rm avail}=300$ L for the $1000$ kg probe, $3$ mL for the $1$ g probe, and $3$ nL for the nanorobot). 

The resulting compression factors are extreme across essentially all payload classes and cruise velocities of interest. For the kilogram-class probe, Figure~2 indicates that $C$ exceeds $\sim 10^{12}$ already at $\beta \sim 0.1$ and rises toward $\gtrsim 10^{15}$ as $\beta\rightarrow 1$. The gram-class probe requires comparably severe compression, remaining above $C\sim 10^{9}$ for relativistic cruise. Even in the nanorobot limit, the required compression factor exceeds $\sim 10^{7}$ for $\beta \gtrsim 0.1$, increasing rapidly with cruise velocity.

Including braking increases $C$ by a factor of two at fixed $\beta$ (dashed curves).

\begin{figure}[h]
    \centering
    \includegraphics[width=\columnwidth]{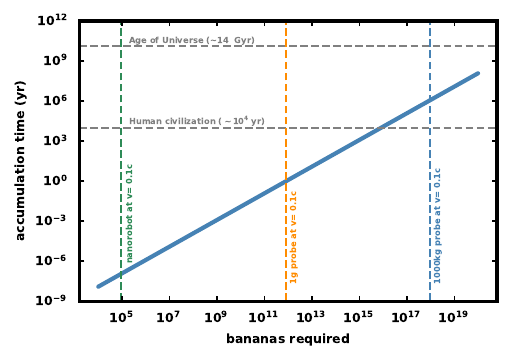}
    \caption{Time taken to produce the number of bananas required. For many scenarios, it is longer than the age of humanity. For a starship capable of carrying a small village, it would exceed the age of the Universe}
    \label{fig:example}
\end{figure}

Figure~3 places the banana requirements in the context of terrestrial resource production. While accumulation times are negligible for nanorobot-scale payloads at low cruise velocities, gram- and kilogram-scale missions require banana inventories that cannot be accumulated on human or even cosmological timescales under present-day production rates. This conclusion holds even when assuming that global banana production can be entirely diverted to propulsion purposes, neglecting all competing demands.

\section{Discussion}
\label{sec:discussion}

We do not provide a discussion to the reader, as this paper is already bananas.

\section{Conclusion}
\label{sec:conclusion}

In this work, we have examined the feasibility of banana-derived positron production as an antimatter source for interstellar propulsion. 

The results show that the banana requirements increase rapidly with cruise velocity and scale linearly with payload mass. High banana demands  translate into extreme compression factors that far exceed any physically plausible concentration of biological matter. Finally, accumulation times, when benchmarked against present-day global banana production rates, exceed human or cosmological timescales for all but the smallest payloads and lowest cruise velocities.

Despite these constraints, bananas remain a  surprisingly viable antimatter source, particularly for low-mass payloads. For nanorobots, banana inventories and accumulation times fall to practical levels, suggesting that biologically sourced positrons could support limited interstellar precursor missions. In this sense, bananas may not solve the problem of interstellar travel, but they do point toward a regime in which it becomes marginally less absurd.

\bibliographystyle{apj}
\bibliography{sample701}

\end{document}